\documentclass{achemso}

\usepackage{graphicx,multirow}

\author{Ersen Mete}\email{emete@balikesir.edu.tr}
\affiliation{Department of Physics, Bal{\i}kesir University, Bal{\i}kesir 10145,
Turkey}
\author{Mehmet Fatih Dan{\i}\c{s}man}\email{danisman@metu.edu.tr}
\affiliation{Department of Chemistry, Middle East Technical University, Ankara
06800, Turkey}

\title[Pentacene Thin Films on Flat and Vicinal Au(111) Surfaces]
{Dispersion Corrected DFT Study of Pentacene Thin Films on Flat and Vicinal Au(111) Surfaces}

\keywords{vdW-DF Calculations, Pentacene, Au Surfaces, Step Edges}

\begin{document}

\begin{abstract}
Here we a present a density functional theory study of pentacene ultra-thin 
films on flat [(111)] and vicinal [(455)] Au surfaces. We have performed 
crystal and electronic structure calculations by using PBE and optB86b-vdW 
functionals and investigated the effects of long range Van der Waals 
interactions for different coverages starting from a single isolated molecule 
up to 4 monolayers of coverage. For an isolated molecule both functionals 
yield the hollow site as the most stable one with bridge-60 site being very 
close in energy in case of optB86b-vdW. Binding strength of an isolated 
pentacene on the step edge was found to be much larger than that on the 
terrace sites. Different experimentally reported monolayer structures were 
compared and the (6 $\times$ 3) unit cell was found to be energetically more 
stable than the ($2 \times 2 \sqrt{7}$) and ($2 \times \sqrt{31}$) ones. 
For one monolayer films while dispersion corrected calculations favored 
flat pentacene molecules on terraces, standard (PBE) calculations either 
found tilted and flat configurations to be energetically similar (on (111) 
surface) or favored the tilted configuration (on (455) surface). PDOS 
calculations performed with optB86b-vdW functional showed larger dispersion 
of molecular orbitals over the Au states for the (455) surface when compared 
with the (111) surface, indicating an enhanced charge carrier transport at 
the pentacene-gold interface in favor of the vicinal surface. Starting with 
the second monolayer, both functionals favored tilted configurations for 
both surfaces. Our results underline the importance of the dispersion 
corrections for the loosely bound systems like pentacene on gold and the role 
played by step edges in determining the multilayer film structure and charge 
transfer at the organic molecule-metal interface.
\end{abstract}

\section{Introduction}

In the last 30 years organic electronics have evolved to a point that 
commercial devices are now a part of our lives in the form of, for example, 
organic light emitting diodes and lightening applications. Nevertheless, 
there are still many issues that need a fundamental understanding and further 
development.\cite{Coropceanu,Klauk,Wang,Facchetti} Pentacene is one of the 
most studied small molecule organic semiconductors and find use in field effect 
transistors.\cite{Wang,Ruiz,Anthony} Among the many factors that affect the 
performance of an OFET are the charge transfer at the metal electrode/organic 
semiconductor interface and the mobility of the charges in organic semiconductor 
film.\cite{Coropceanu,Klauk,Wang,Facchetti,Ruiz,Anthony} Both of these 
factors are in turn, directly or indirectly, correlated to the structure and morphology 
of the organic film on the metal electrode to a certain extent. Hence electronic 
and structural properties of pentacene films on metal surfaces have been 
investigated intensively both experimentally\cite{Casalis,Danisman,Albayrak,France1,France2,Schroeder1,France3,Kang1,Kang2,Kafer,Pedio,Dougherty,Duhm} 
and theoretically.\cite{Muller,Pieczyrak,Li,Toyoda,Wheeler,Schatschneider,Mete1,Mete2,Lee1,Ferretti,Baldacchini,Simeoni,Sun,Saranya,Cantrell,Lee2} 

\begin{figure}
\includegraphics[width=8.3cm]{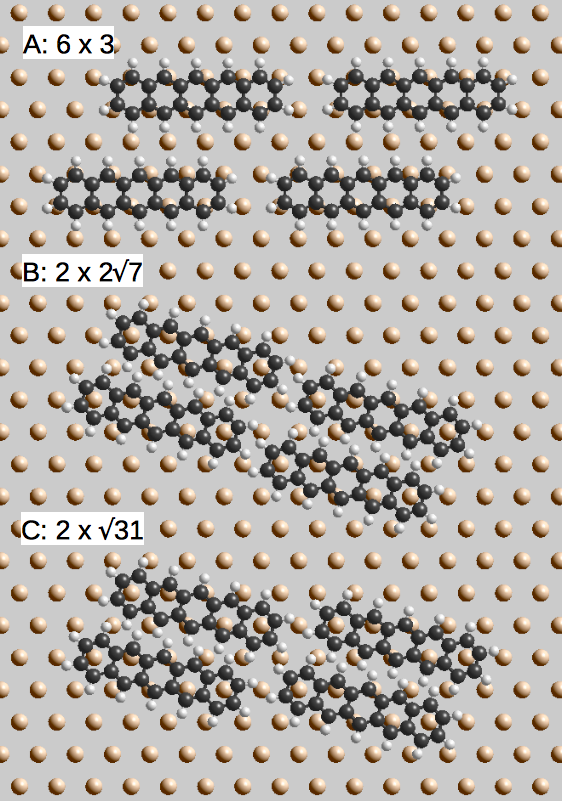}
\caption{Structural models of pentacene full coverage monolayers on Au(111) 
surface. The molecular positions are not meant to depict the actual adsorption 
sites. The molecules are drawn perfectly parallel to the surface (without any 
tilt).\label{fig1}}
\end{figure}

Since gold is the choice of electrode material in many devices, pentacene on 
Au surfaces have attracted particular interest.\cite{Albayrak,France2,Schroeder1,France3,Kang1,Kang2,Kafer,Pieczyrak,Toyoda,Wheeler,Lee1,Lee2} 
By means of scanning tunneling microscopy (STM)\cite{France1,France2,Kang1,Kang2,Kafer} 
and x-ray absorption spectroscopy (NEXAFS)\cite{Kafer} several groups have reported 
similar  full coverage monolayer structures with pentacene molecules lying flat on 
the Au (111) surface if not with a small tilt angle (NEXAFS results indicate 
an average tilt angle of 13$^\circ$ \cite{Kafer}). These structures which will be referred 
to as ``B'' and ``C'' are shown in Fig.~\ref{fig1}  and have packing density of about 
1.1 x 10$^{14}$ molecules/cm$^2$. Recently we have performed a helium diffraction 
study of the Pen/Au(111)\cite{Albayrak} system and observed a commensurate 
6 $\times$ 3 monolayer structure which we had also observed on Ag(111) 
surfaces\cite{Casalis,Danisman} This monolayer phase which has a slightly lower packing 
density of 0.8 $\times$ 10$^{14}$ molecules/cm$^2$ is also shown in Fig.~\ref{fig1} and will be 
referred to as ``A''. At higher coverages (starting with 2nd ML) Pen molecules still had their 
long axis parallel to surface but with a higher tilt angle resulting in multilayers structures 
similar to b-c face of the bulk Pen crystal. 

Recently several density functional theory (DFT) studies of Pen/Au(111) have 
been performed focusing mostly on the electronic properties such as work 
function change of the Au surface upon pentacene adsorption.\cite{Pieczyrak,Li,Toyoda,Wheeler}
In these studies either isolated pen molecules or one monolayer films were 
considered on Au(111) surface\cite{Pieczyrak,Li,Toyoda} or on Au clusters\cite{Wheeler},  
without a systematic search for the best adsorption site/geometry and/or the crystal 
structure. Instead the electronic calculations were performed by using a 
chosen pentacene unit cell which were different then the experimentally 
determined ones described above. Nevertheless the calculated work function 
changes were in good agreement with the experimental value (0.95 eV~\cite{Schroeder2}) 
regardless of the functionals, methods or the unit cells used when the 
calculations were made on the Au(111) surfaces. In the cluster calculations 
of Dahnovsky et al.\cite{Wheeler}, however, who studied the potential energy 
surface of pentacene and found a very a smooth landscape, the calculated work 
function change was much lower than the experimental value.  Long range van der 
Waals interactions were considered in three of these studies with 
different approaches. Morikawa et al.\cite{Toyoda} employed DFT-D\cite{Grimme} 
and vdW-DF\cite{Dion}methods and found that while the former method well 
reproduces the Pen-substrate distance, the latter method well reproduces the binding 
energies. Dahnovsky et al. on the other hand employed CAM-B3LYP\cite{Yanai}, 
wB97 and wB97X\cite{Chai} long range correlated functionals and found the latter 
two functionals to reproduce the experimental binding energies accurately. Finally, 
Ortega et al.\cite{Pieczyrak} who employed LCAO-S2-vdW\cite{Dappe} formalism 
for geometry optimization could achieve electronic structure results comparable to the 
experimental values. 

Here we have systematically studied structural and electronic properties of 
pentacene on flat and vicinal Au(111) surface as a function of coverage by 
means of van der Walls density functional theory (vdW-DF) calculations. We 
first studied isolated molecules to determine the most preferred adsorption site 
and the potential energy surface. Then we have examined different monolayer 
structures which were reported in the literature. Finally we have investigated 
higher coverage films up to four monolayers. By this way we were able to 
monitor the evolution of the electronic and structural properties of the pen films 
as a function of film thickness and unit cell structure.

\section{Theoretical Methods}

In order to describe the morphology and the electronic structures of Pen 
multilayers on flat and vicinal Au(111) surfaces, we performed van der Waals 
(vdW) density functional theory (DFT) calculations based on projector augmented
waves (PAW)\cite{Blochl} method available in VASP.\cite{Kresse1,Kresse2,Klimes} 
A self-consistent implementation of the vdW-DF method of 
Dion \textit{et al.}\cite{Dion} was used as a non-local correlation functional 
which accounts for the dispersion interactions. The exchange part was
treated with optB86b functional which is optimized for the correlation part.
For the prediction of material properties, the optB86b-vdW approach has been 
recently shown to improve over standard and revised exchange-correlation (XC) 
schemes like PBE, revPBE and PBEsol.\cite{Klimes}

\begin{figure}
\includegraphics[width=8.4cm]{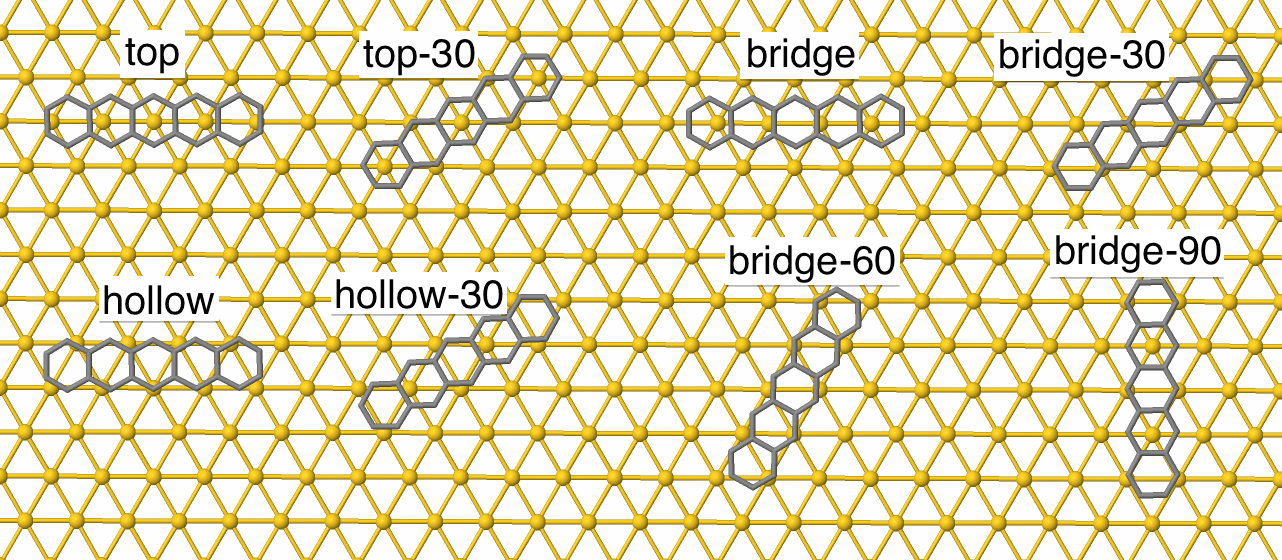} 
\caption{Possible adsorption sites of an isolated pentacene molecule on the 
flat Au(111) surface.\label{fig2}}
\end{figure}

Four layer slab models were constructed from the bulk phase of gold to 
represent the flat and vicinal Au(111) surfaces. The ionic positions were 
optimized by minimizing the Hellmann-Feynman forces until a threshold value 
of 0.01 eV/{\AA} was reached on each atom. After a full relaxation of the slab 
geometries, the displacement of the atoms from their bulk positions remained
to be less than $10^{-3}$ {\AA}. Therefore the slab thickness is found to be
sufficient to represent Au(111) surface properties. Then, in order to speed up
the calculations we only kept the Au atoms at the bottom layer of the slabs 
frozen to their bulk positions. Isolated Pen molecules were considered on the 
5$\times$8 flat slab to accommodate sufficiently large separation between 
their periodic images. All possible adsorption configurations were taken into 
account as shown in Fig.~\ref{fig2}. Single and multi layer Pen coverages were 
modeled on the flat Au(111)-3$\times$6 supercell. We considered different Pen 
monolayer phases as seen in Fig.\ref{fig1} Several probable adsorption 
configurations were built for the Pen mono- and multilayers similar to the 
isolated case. For the vicinal Au(111) surface, an Au(455) supercell was built 
with a terrace size of $\sim$19.8 {\AA} along [$\bar{2}$11]. A vacuum space 
with a height of least 13 {\AA} was introduced to prevent any unphysical 
interaction between the periodic images of the slabs. All calculations were 
performed both at the standard DFT level with PBE XC functional and at the 
vdW-DF level with optB86b-vdW approach.

The interactons between the ionic cores and the valence electrons have been 
treated within the PAW method using plane waves up to a cutoff energy of 
370 eV. Brillouin zone sampling was performed on a $k$-point mesh of 
4$\times$4$\times$1. The density of states (DOS) calculations have
been carried out with doubly denser $k$-point meshes. These parameters were 
tested and optimized to get well converged total energies of the physical 
structures considered in this work.

We estimated the binding energies of Pen molecules on Au(111) surface by,
\[
E_{\rm b}=E_{{\rm Pn}/{\rm Au(111)}}-E_{\rm Au(111)}-E_{\rm Pn}\, ,
\]
where $E_{{\rm Pn}/{\rm Au(111)}}$, $E_{\rm Au(111)}$ and $E_{\rm Pn}$ are 
the total energies of the Pn/Au(111) combined system, of the bare Au(111) slab 
and of a single Pn in a big box, respectively. 

\section{Results and Discussion}

Gold forms in a \textit{ccp} crystal structure having a space group symmetry 
of Fm$\bar{3}$m (225) with a bulk lattice constant of 4.078 {\AA}.\cite{Wyckhoff} 
Standard PBE overestimates it leading to a value of 4.160 {\AA} due to the 
inherent tendency of the LDA to distribute charge uniformly which is not the 
case especially in metals. The vdW-DF method better predicts the lattice 
constant to be 4.125 {\AA} through the formulation of the long ranged 
dispersive effects the correlation part and the optimization of exchange 
energy, so called the optB86b functional. 

\begin{table}[htb]
\caption{Relative total cell energies $E_{\rm t}$ and adsorption heights $h$ 
of a single Pen molecule at different sites on Au(111) surface calculated with 
PBE and optB86b-vdW methods.
\label{table1}}
\begin{tabular}{lcccc}\hline
\multirow{2}{*}{Site} & \multicolumn{2}{c}{PBE} & \multicolumn{2}{c}{optB86b-vdW}\\
& E$_{\rm t}$ {\footnotesize(eV)} & 
$h$ {\footnotesize({\AA})} & 
E$_{\rm t}$ {\footnotesize(eV)} & 
$h$ {\footnotesize({\AA})} \\[1mm]\hline
top & 0.025 & 4.06 & 0.033 & 3.25\\
top-30 & 0.035 & 4.07 & 0.201 & 3.08\\
bridge & 0.211 & 4.09 & 0.243 & 3.14\\
bridge-30 & 0.213 & 4.08 & 0.176 & 2.96\\
bridge-60 & 0.190  & 4.07 & 0.007 & 2.94\\
bridge-90 & 0.206 & 4.09 & 0.106 & 3.10\\
hollow & 0.000 & 3.87 & 0.000 & 2.94\\
hollow-30 & 0.029 & 4.00 & 0.145 & 3.12\\ \hline
\end{tabular}
\end{table}

\textbf{Isolated Pentacene on Au(111).}
The possible adsorption sites and orientations of a single pentacene on the 
flat Au(111) surface are presented in Figure~\ref{fig2}. A Pen molecule can be
considered as isolated on a sufficiently large (8 $\times$ 5) Au surface cell 
on which the separation between its periodic images is at least 8 {\AA}. Each 
of the possible initial geometries were relaxed including additional 
standing-up configurations which are energetically 0.5 eV (vdW included) less 
favorable relative to the planar cases. Both the standard and vdW DFT 
calculations find the hollow configuration as the minimum energy adsorption 
site as seen in Table.~\ref{table1}. The relative total energies between 
different adsorption orientations get as large as $\sim$0.2 eV per molecule 
suggesting a rather shallow plateau-like potential energy surface (PES) for 
the flat Au(111) termination. France \textit{et al.} determined the adsorption 
energy as 110 kJ/mol from initial monolayer of Pen on  the Au(111) surface
by performing temperature-programmed desorption (TPD) 
experiments.\cite{France2} The standard PBE significantly underestimates 
the binding energy of a single Pen on the flat Au surface. Its prediction is 
a weak adsorption at the hollow site with an energy of 0.332 eV (32 kJ/mol) 
at a height of 3.87 {\AA}. Moreover, the penalty in the energy from the 
planar to a slightly tilted position is negligibly small with the PBE 
functional. However, the relaxation with the vdW-DF method gives a binding 
energy of 1.337 eV (129 kJ/mol) per molecule preferring the planar adsorption 
at a relatively shorter height of 2.94 {\AA} at the hollow site. Our vdW-DF
calculations reveal that the bridge-60 and the hollow configurations are 
energetically similar. Pen molecules at the both adsorption orientations 
follow the lattice symmetry similarly. This indicates that the matching 
between the molecular and surface charge densities is an important factor in 
the determination of the preferential adsorption site. In fact in our previous 
DFT calculations for Pen on Ag(111), bridge 60 and hollow configurations were 
also found to be very close in energy, bridge 60 being slightly more stable 
(by 8 meV).\cite{Mete1}

\begin{figure}
\includegraphics[width=8.4cm]{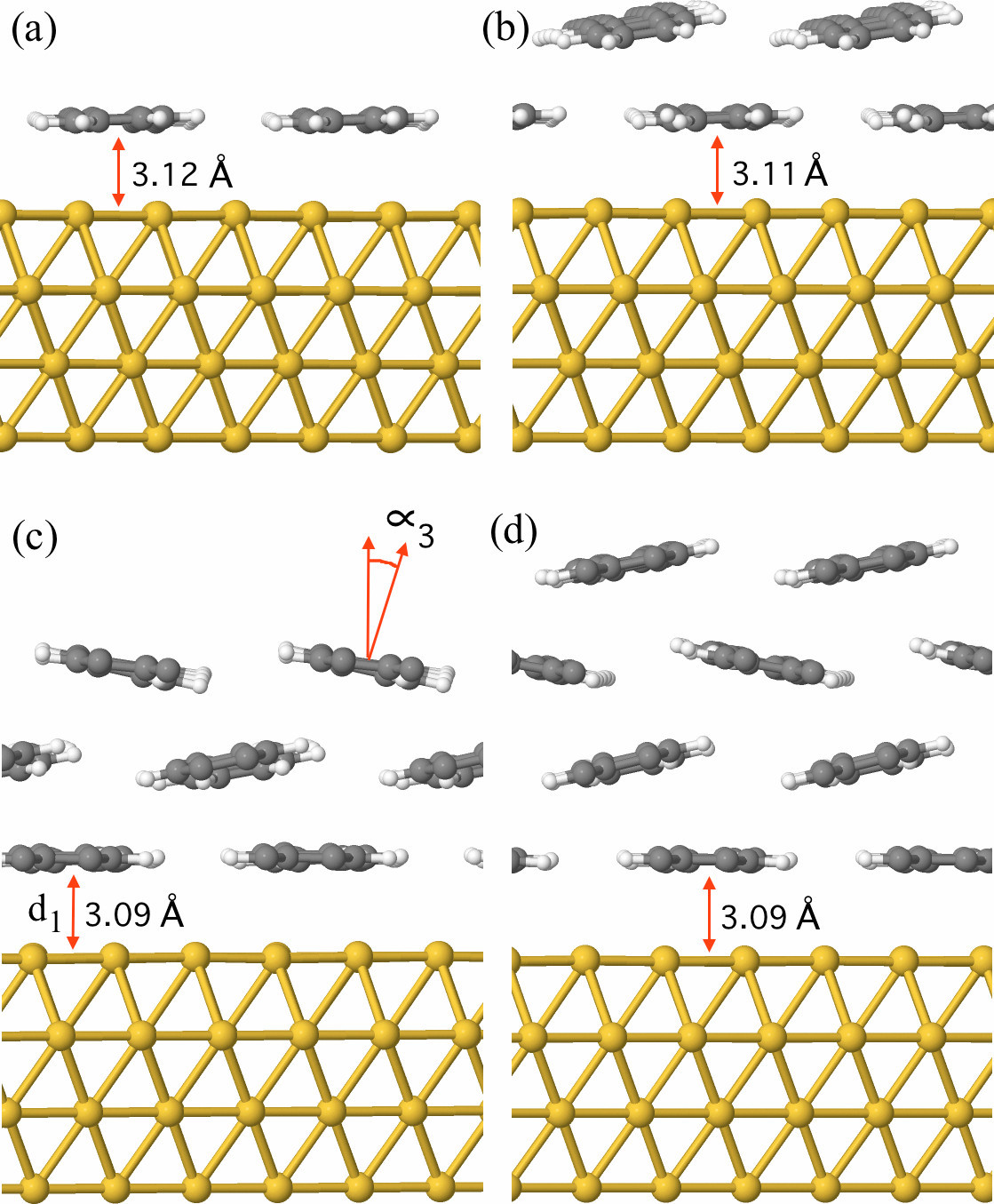} 
\caption{The atomistic structure of pentacene multilayers on the flat 
Au(111) surface optimized with optB86b-vdW method.\label{fig3}}
\end{figure}

The small energy differences between different adsorption orientations in 
Table~\ref{table1} indicate a rather flat potential energy surface (PES) which 
could allow the diffusion of the Pen molecules at the contact layer. For 
instance, physisorbed Pen molecules were reported to be dragged by the 
STM tip during image acquisition on the Au(111) surface.\cite{Soe}

\begin{table}[htb]
\caption{Energetics and structural parameters calculated with PBE 
for Pen on flat Au(111) surface for different coverage models. 
$E_{\rm t}$ are the relative supercell total energies in eV. Average interlayer 
separations, $d$ (in {\AA}), and tilting angles, $\alpha$ (in degrees), are 
labeled with corresponding layer numbers. Angular deviations of the molecules 
from the [1$\bar{1}$0] surface symmetry line are presented in parantheses for 
the corresponding layer.
\label{table2}}
\footnotesize
\begin{tabular}{llccccccccc}
Film & Site & E$_{\rm t}$ & $d_1$
&$d_{12}$ & $d_{23}$ &$d_{34}$ & $\alpha_1$ & $\alpha_2$ & $\alpha_3$ & $\alpha_4$
\\[1mm]\hline
\multirow{3}{*}{1 ML} & top & 0.022 & 3.98 &&&& 0.0(0.5) &&& \\
& bridge-60 & 0.000 & 3.98 &&&& 0.0(0.8) &&&\\
& hollow & 0.003 & 3.99 &&&& 0.0(0.5) &&&\\[1mm] \hline
\multirow{4}{*}{2 ML} & flat@bridge-60\hspace{-2mm} & 0.061 & 3.94 & 3.64 &&& ~0.0(1.4) & ~0.0(1.3) &&  \\
& top & 0.062 & 3.91 & 3.51 &&& ~0.0(3.0) & 21.7(4.7) &&\\
& bridge-60 & 0.032 & 3.67 & 3.67 &&& -3.5(2.8) & 20.5(5.0) &&\\
& hollow & 0.000 & 3.67 & 3.65 &&& -10.3(2.9) & 20.3(5.0) &&\\[1mm] \hline
\multirow{3}{*}{3ML} & top & 0.002 & 3.81 & 3.81 & 3.81 && -7.5(5.8) & 24.4(4.7) & -24.9(3.4)&\\
& bridge-60 & 0.000 & 3.81 & 3.46 & 3.71 && -5.4(1.7) & 25.1(6.3) & -22.9(3.2) &\\
& hollow & 0.032 & 3.81 & 3.46 & 3.54 && -7.6(3.8) & 22.2(5.6) & -25.1(3.9) &\\[1mm] \hline
\multirow{3}{*}{4ML} & top & 0.001 & 3.78 & 3.11 & 3.46 & 3.81 & -10.3(8.5) & 25.2(9.6) & -26.5(5.1) & 28.2(4.5)\\
& bridge-60 & 0.000 & 3.42 & 3.52 & 3.63 & 3.64 & -6.2(4.1) & 25.4(5.6) & -24.7(3.7) & 21.6(3.7)\\
& hollow & 0.087 & 3.46 & 3.17 & 3.41 & 3.81 & -4.9(3.7) & 21.5(8.9) & -26.1(4.6) & 22.2(3.3) \\ \hline
\end{tabular}
\end{table}

\begin{table}[htb]
\caption{Energetics and structural parameters calculated with optB86b-vdW 
for Pen on flat Au(111) surface for different coverage models. $E_{\rm t}$ are 
the relative supercell total energies in eV. Average interlayer separations, 
$d$ (in {\AA}), and tilting angles, $\alpha$ (in dgrees), are labeled with 
corresponding layer numbers.
\label{table3}}
\footnotesize
\begin{tabular}{llccccccccc}
Film & Site & E$_{\rm t}$ & $d_1$ & $d_{1\textrm{-}2}$ & $d_{2\textrm{-}3}$ 
& $d_{3\textrm{-}4}$ & $\alpha_1$ & $\alpha_2$ & $\alpha_3$ & $\alpha_4$
\\[1mm]\hline
\multirow{3}{*}{1 ML} & top & 0.183 & 3.25 &&&& ~0.0 &&& \\
& bridge-60 & 0.000 & 3.12 &&&& ~0.0 &&&\\
& hollow & 0.008 & 3.13 &&&& ~0.0 &&&\\[1mm] \hline
\multirow{4}{*}{2 ML} & flat@bridge-60 &  0.070 & 3.13 & 3.11 &&& 0.0(0.0) & ~0.0(0.0) &&  \\
& top & 0.116 & 3.13 & 2.95 &&& 0.0(6.3) & 13.7(4.7) &&\\
& bridge-60 & 0.000 & 3.11 & 2.90 &&& 0.0(2.0) & 12.4(4.7) &&\\
& hollow & 0.014 & 3.12 & 2.93 &&& 0.0(2.0) & 12.2(5.3) &&\\[1mm] \hline
\multirow{3}{*}{3ML} & top & 0.108 & 3.12 & 3.07 & 3.12 && 0.0(7.5) & 10.0(2.9) & -9.9(-2.0)&\\
& bridge-60 & 0.003 & 3.11 & 3.07 & 3.12 && 0.0(2.0) & 10.4(2.3) & -9.5 (-1.3)&\\
& hollow & 0.000 & 3.09 & 3.05 & 3.12 && 0.0(2.1) & 13.1(3.0) & -12.0(-1.5) &\\[1mm] \hline
\multirow{2}{*}{4ML} & bridge-60 & 0.016 & 3.10 & 3.11 & 3.12 & 3.12 & 0.0(1.9) & 13.9(2.0) & -13.4(-2.4) & 10.1(-1.4)\\
& hollow & 0.000 & 3.09 & 3.10 & 3.10 & 3.12 & 0.0(1.9) & 13.7(0.6) & -12.4(-4.5) & 11.1(-2.4) \\ \hline
\end{tabular}
\end{table}

\textbf{Monolayer on Flat Au(111).}
A full Pen monolayer coverage has been modeled on an experimentally 
observed 6$\times$3\cite{Albayrak} Au(111) surface unit cell, referred as the 
phase A in Figure~\ref{fig1}. We considered a number of probable 
configurations including top, bridge-60, and hollow sites for the molecules on 
this contact layer. We optimized the initial geometries by using both the 
standard PBE and the optB86b-vdW methods. The minimum energy adsorption leads 
to the bridge-60 position which is only slightly preferable over the hollow 
site as presented in Table~\ref{table2} and Table~\ref{table3}. The 
computational cell structure of the phase A is not commensurate with those 
of the phases B and C as seen in Figure~\ref{fig1}. To compare their
energetics and interpret these results for the relative stabilities, we 
calculated the binding energies of the full monolayer in each case.
The corresponding values are obtained as 0.23 eV, 0.17 eV , and 0.15 eV 
for the phases A, B and C, respectively. Hence, the adsorption of the 
contact layer is energetically more preferable with the 6$\times$3 cell 
structure on the Au(111) surface. Nevertheless, all these three configurations 
are very close in energy which explains why pentacene has so many different 
monolayer phases/structures that form simultaneously on the Au(111) surface.
Due to in-plane intermolecular interactions, the binding energy in the full 
monolayer coverage substantially drops relative to the single isolated 
adsorption case. The standard PBE calculations predict much weaker Pen-Au 
interaction compared with the vdW-DF results. For instance, the adsorption 
height of the full monolayer is 3.98 {\AA} and 3.12 {\AA} with PBE and 
optB86b-vdW methods, respectively. In addition, when started from sightly 
tilted molecular configuration, PBE tends to keep the geometry with a 
negligibly small increase in the total energies while optB86b-vdW prefers 
the molecules to become flat over the Au(111) surface, still portraying 
a weak metal-molecule interaction. Previous theoretical studies found 
Pen-Au distance as 3.2 {\AA} by including weak London dispersion forces 
semiempirically.\cite{Pieczyrak,Toyoda} Therefore, the optB86b-vdW method 
gives a relatively better description of the adsorption characteristics of 
the Pen molecules on the flat Au(111) surface. We presented the atomistic 
structures of the adsorption geometries of the multilayer Pen coverages in 
Figure~\ref{fig3}.  

Our standard and vdW corrected DFT calculations suggest a flat 1 ML 
physisorption instead of a tilted chemisorption of a full Pen layer.  In 
fact, the planar monolayer was reported by experiments.\cite{Kafer,Kang2} 
based on their NEXAFS and thermal desorption signatures. The weakness of 
Pen-Au interaction reproduced through our density functional analyses does 
not allow to rule out the possibility of an average tilt at the contact layer 
which may develop as a consequence of the experimental conditions. Our vdW 
corrected DFT calculations favor the molecules in the monolayer (ML) to follow 
the surface symmetry and to be aligned almost parallel to the gold rows. 
However, the standard PBE XC functional leads to a slight deviation from 
the [1$\bar{1}$0] symmetry line with a small angle ($\sim$0.5$^\circ$) as 
indicated in the parantheses in Table~\ref{table2}.

\textbf{Thin Films on Flat Au(111).}
For the second Pen ML, we considered flat-lying and tilted molecules at different 
adsorption sites with and without long range correlations as presented in 
Table~\ref{table2} and Table~\ref{table3}. In the forms of two-layer structure, 
the contact layer molecules energetically prefer the hollow and the bridge-60 
adsorption sites at the PBE and optB86b-vdW levels, respectively. The second 
ML becomes tilted as shown in Figure~\ref{fig3}b. The molecular tilting angles 
are calculated around 3$^\circ$ to 4$^\circ$, and are in agreement with 
previous experiments.\cite{France2} The PBE XC functional yields a large 
separation of 3.7 {\AA} between the contact and the second ML while a moderate 
value of 3.1 {\AA} was predicted by the optB86b-vdW method. Similarly,
the tilting angles of Pen molecules are significantly larger when dispersive
forces are omitted. The possibility of flat-lying upper layers has been considered 
for the multilayers. Those initial structures relaxed to tilted pentacenes above the 
planar contact layer in the 3 and 4 ML cases. A fully flat 2 ML configuration at the 
bridge-60 site is energetically less favorable (by $\sim$70 meV at the vdW-DF level) 
relative to the minimum energy geometry. The total energies of the bridge-60 and 
hollow sites are very close to each other since the molecular charge densities  
of Pens at the interface have similar distributions with respect to surface gold 
network.  The small energy barriers between different adsorption cases also point 
out the weakness of molecule-metal interaction mimicking the energy differences 
between similar Pen film phases.

The interlayer distances are overestimated by PBE calculations while vdW-DF 
reasonably converges them to $\sim$3.1 {\AA} as seen in Figure~\ref{fig3}. 
In the case of multilayers, the molecular axes deviate from the surface gold 
rows. Standard DFT calculations give larger deviations inconsistent with 
experiments. The inclusion of the long-range correlations in the form of a 
self-consistent vdW-DF formulation become important to get a sound 
description of both Pen-Au and Pen-Pen interactions. As the number of layers 
increases, intermolecular interactions causes pentacene molecules to adopt a 
thin film phase on the flat gold surface. 

\begin{figure}
\includegraphics[angle=-90,width=8.4cm]{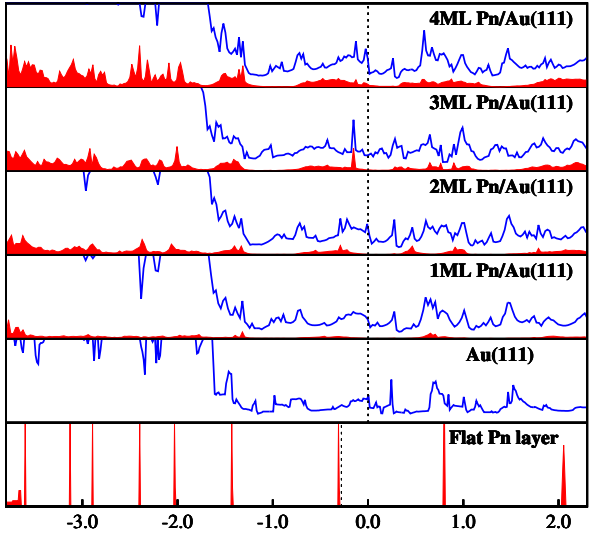} 
\caption{The optB86b-vdW results for the total DOS of Pen/Au(111) structures 
are drawn with blue lines. The projected DOS indicate the pentacene 
contributions as the red shades. The energy scale is given in eV and its zero 
is chosen relative to the energy of the highest occupied state of flat Au(111) 
surface. The dotted lines denote the Fermi energies separately for each 
case.\label{fig4}}
\end{figure}

\textbf{Electronic Structures of Pentacene Films on Flat Au(111).}
We calculated the densities of states (DOS) using optB86b-vdW
method and presented the results in Figure~\ref{fig4}. The bottom 
panel shows the calculated DOS of 1 ML Pen without the Au(111)
slab. In-plane Pens have very well localized states with negligibly 
small dispersion in the form of sharp peaks due to weak overlap of 
the molecular orbitals. The Fermi level is chosen as the energy of 
the highest occupied state. The zero of the energy scale is determined
with respect to Fermi energy of the bare Au(111) surface. The HOMO
and the LUMO levels of Pen are identified as -0.36 eV and 0.79 eV
relative to the Fermi energy.  Therefore, the band gap of Pen is
predicted as 1.15 eV which smaller than the experimental value 
of 1.85 eV.\cite{Schroeder1,Schroeder2} This is a well-known 
underestimation of DFT due to the lack of proper cancellation of
self-interaction between the Hartree and the exchange terms. In
other words, standard XC functionals are not formulated to grasp
the excitation processes in materials. Therefore, the unoccupied
states are not properly described. This also applies to the vdW
corrected DFT.  

The DOS plot of 1 ML Pen on Au(111) is shown in the third panel 
from the bottom in Figure~\ref{fig4}. The shades corresponding to
the Pen contribution to the electronic structure of the combined 
system show almost no shift in energy when compared with
the flat Pen layer in the absence of the gold slab. The weak
coupling between the Pen contact layer and the gold surface 
is emphasized by the broadening of the molecular energy levels
of Pen molecules over a small number of gold states. 
Therefore, the PDOS of the contact layer exhibits the characteristics
of a film phase with its frontier molecular states contributing
to the electronic structure of the combined system around the 
Fermi energy. 

In the case of multilayers, PDOS satellites form and localize 
at around the flat Pn layer peak positions. Therefore, vdW-DF 
calculations indicate a relatively stronger molecular orbital overlap
between the in-plane pentacenes in the same layer. The broadening 
of these satellite structures gets larger with increasing number of 
full monolayers. One of the main factors is associated with the long 
range correlation effects and is coming from the coupling between 
the molecules at different layers. In particular, the Pen contribution 
in the case of 4ML Pn/Au(111) system reflect a thin film formation.
The positions of the highest occupied and lowest unoccupied states of 
$\pi$-stacked pentacenes with respect to the Fermi energy of the metal 
surface show no electronic band transport feature. However, the 
multilayer structures organize in a well ordered thin film phase leading
to a large $pi$-conjugation length along the molecular axis due to the 
overlap of the molecular orbitals between the nearest neighbor 
pentacenes. Therefore, the charge carrier transport can be explained as
a result of a hopping mechanism between the localized Pen states.

\begin{figure}
\includegraphics[width=8.4cm]{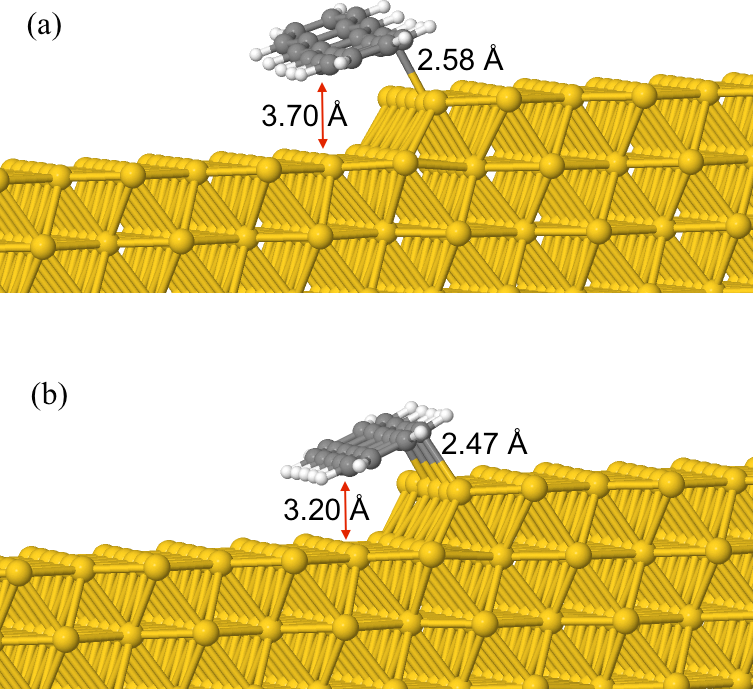} 
\caption{The minimum energy adsorption structure of an 
isolated Pen molecule at the step edge on the Ag(455) surface 
optimized with a) PBE and b) optB86b-vdW methods.\label{fig5}}
\end{figure}

\begin{figure}
\includegraphics[width=8.4cm]{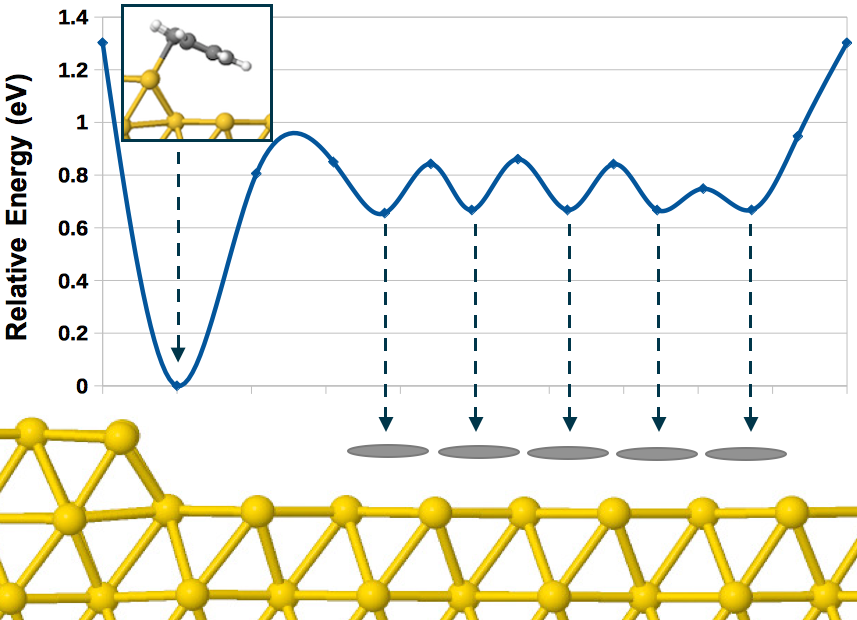} 
\caption{Potential energy profile of single Pen molecule on the
Ag(455) vicinal surface along the [$\bar{2}$11] direction. 
 \label{fig6}}
\end{figure}

\textbf{Isolated Pentacene on Vicinal Gold.}
We considered an isolated Pen molecule on Au(455) cell on the terrace 
and at the step edge with probable adsorption configurations. The 
initial geometries on the terrace were optimized using both PBE and 
optB86b-vdW methods. The results come out almost identical to those of 
flat Au(111) surface as tabulated in Table~\ref{table1}. Therefore, the 
low energy adsorption on the terrace turns out to be the hollow site. 
Then, we performed structural optimization of a single Pen molecule at 
the step edge using both the standard DFT and the vdW-DF calculations. 
The molecule at the step edge gets tilted about its major axis with 
respect to the Au(111) plane normal as shown in Figure~\ref{fig5}.
The tilting angle is calculated as 20.6$^\circ$ and 26.4$^\circ$ 
with PBE and opt86b-vdW exchange-correlation schemes, respectively.
The adsorption of a single Pen at the step edge is remarkably stronger 
in comparison to that on the flat terrace. Then, we obtained the 
diffusion barrier profile of a Pen along [$\bar{2}$11] direction 
using the self-consistent optB86b-vdW approach. The plot in 
Figure~\ref{fig6} has been drawn by calculating the total energy of
the system where the adsorbate molecule is placed at different locations 
along the line perpendicular to the step row on the Au(455) slab. The 
minimum of the potential occurs for the tilted@step geometry. The energy
is  0.67 eV  lower than flat@hollow adsorption at the optB86b-vdW 
level of theory. This difference turns out to be only 0.32 eV using the PBE 
XC functional as presented in Table~\ref{table4}. The local minima over 
the terrace are due to the adsorption at the hollow site. The barrier heights 
between two successive hollow sites correspond to the bridge-60 
configuration with a value of $\sim$0.19 eV.

The relaxation of an isolated Pen at the step with the two methods brings 
some important structural differences as seen in Figure~\ref{fig5}.
When we use the standard PBE XC functional, the shortest bond between 
the molecule and an Au atom at the step edge is found to be 2.58 {\AA} 
at a height of 3.70 {\AA} from the surface plane with a significant convex 
bending. However, vdW-DF approach leads to even shorter and four bonds 
almost equal in length (2.47 {\AA}). This time, Pen stays 3.20 {\AA}
above the gold surface and is perfectly planar. The binding energy of a 
single Pen molecule at the step edge is calculated using the PBE XC 
functional and the optB86b-vdW method as 0.65 eV (63 kJ/mol) and 
2.28 eV (220 kJ/mol), respectively. The inclusion of weak dispersive 
interactions cause a remarkable difference in the adsorption characteristics 
of Pen molecules on the vicinal gold surfaces.

\begin{table}[htb]
\caption{Relative total cell energies $E_{\rm t}$ and average adsorption 
heights $d_1$ of various Pen coverage models on the vicinal Au(455) 
surface calculated with PBE and optB86b-vdW methods. The bond lengths 
of a single molecule adsorbed at the step are given in parantheses. 
\label{table4}}
\begin{tabular}{ll|cc|cc}\hline
\multirow{2}{*}{Coverage} &\multirow{2}{*}{Model} 
& \multicolumn{2}{c}{PBE} & \multicolumn{2}{c}{optB86b-vdW}\\ 
& & E$_{\rm t}$ {\footnotesize(eV)} & 
$d_1$ {\footnotesize({\AA})} & 
E$_{\rm t}$ {\footnotesize(eV)} & 
$d_1$ {\footnotesize({\AA})} \\[1mm]\hline
\multirow{2}{*}{single} & flat@hollow & 0.322 & 3.87 & 0.668 & 3.12\\
& tilted@step & 0.000 & 3.70 (2.58) & 0.000 & 3.20 (2.47)\\ \hline
\multirow{3}{*}{1 ML} &flat & 0.370 & 3.99 & 0.755 & 3.20 \\
& flat2 & 0.083 & 3.96 & 0.000 & 3.18 \\
& tilted & 0.000 & 3.92 & $\longrightarrow$ & ``flat2" \\ \hline
\multirow{3}{*}{2 ML} &flat-tilted & 0.374 & 3.79 & 1.266 & 3.10 \\
& flat2-tilted & $\longrightarrow$ & ``tilted-tilted" & 0.000 & 3.06 \\ 
& tilted-tilted & 0.000 & 3.29 & $\longrightarrow$ & ``flat2-tilted" \\ \hline
\end{tabular}
\end{table}

\textbf{Pentacene Multilayers on Vicinal Au(455).}
Several initial configurations have been considered to optimize 
a full Pen monolayer coverage on the vicinal surface using both
the standard and vdW corrected DFT.  These involve tilted, flat 
and flat2 cases. The tilted case corresponds to the model where 
all molecules are tilted with respect to the surface normal about 
their major axes.  In the flat one, every Pen lays parallel over the 
surface as that on the planar Au(111) case. The flat2 model was 
constructed such that the molecules lie  planar on the terrace and 
tilted at the step edge. 

\begin{figure}
\includegraphics[width=8.4cm]{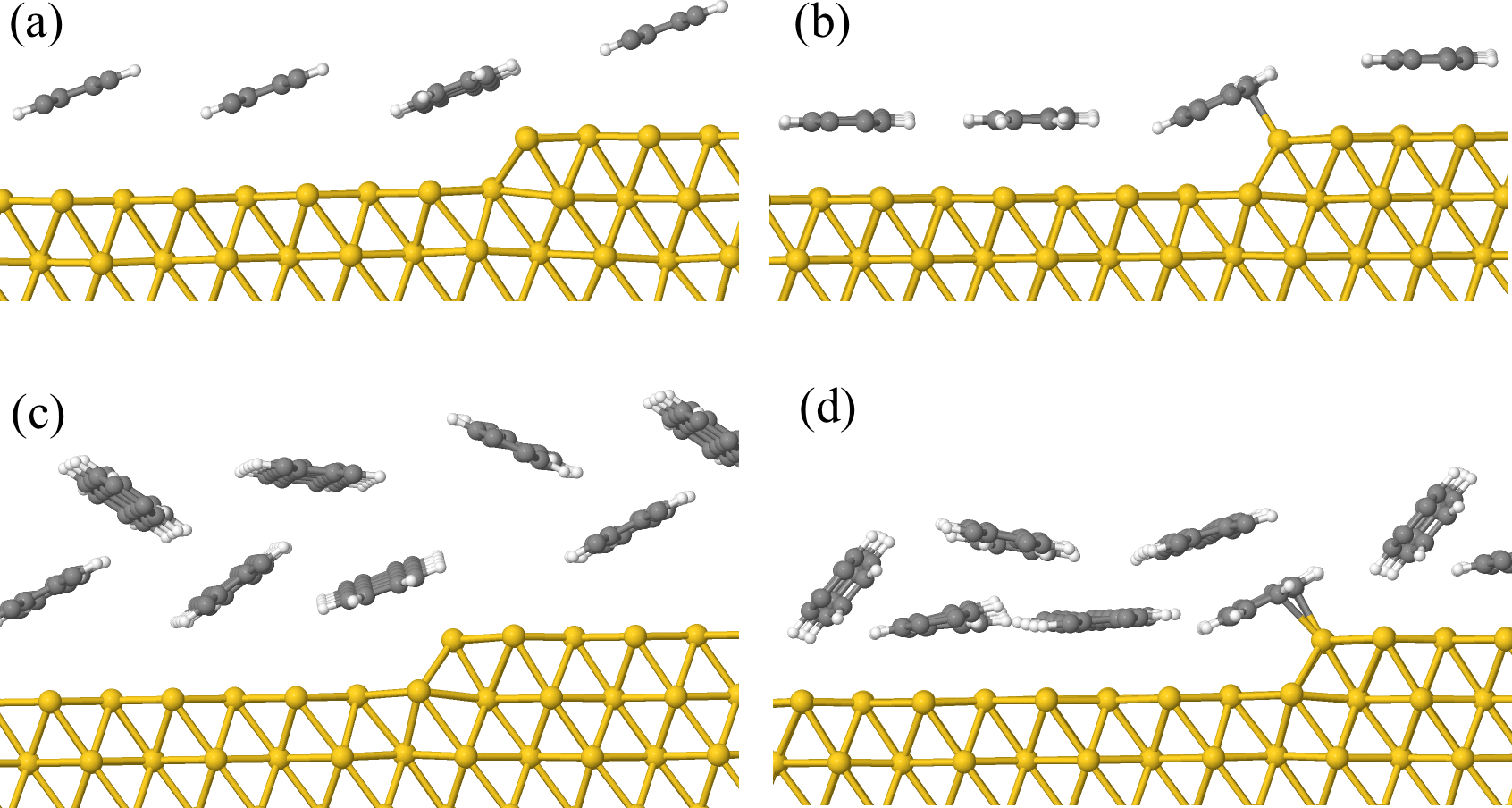} 
\caption{The minimum energy geometries of one and two Pen monolayers on 
the Ag(455) surface optimized with PBE and optB86b-vdW methods. 
 \label{fig7}}
\end{figure}

The minimum energy geometry of a full Pen monolayer on the 
vicinal Au surface with the standard DFT (PBE) turns out to be 
the tilted configuration as shown in Figure~\ref{fig7}a. The tilting
angle is about 22.8$^\circ$. Apparently, this result is different 
from the case of 1 ML on the flat Au(111) surface. Although the 
binding energy of a tilted Pen at the step edge is underestimated 
by PBE, the molecules on the terrace follow this tilting since 
intermolecular coupling is relatively stronger than the 
metal-molecule interaction at this level of theory. However, the 
vdW-DF calculations (optB86b-vdW) favor the flat2 case 
(Figure~\ref{fig7}b) as a result of the inclusion of the dispersive 
forces self-consistently. This improves the description of organic 
molecule-metal surface interaction. For instance, when started from 
the tilted model as the initial geometry, the vdW calculations ends 
up with the flat2 configuration. Moreover, as presented in 
Table~\ref{table4}, the height of the adlayer from the terrace gets 
slightly larger than the separation of a single isolated molecule from 
the surface, mimicking the role of molecular orbital overlap. 
 
\begin{figure}
\includegraphics[angle=-90,width=8.4cm]{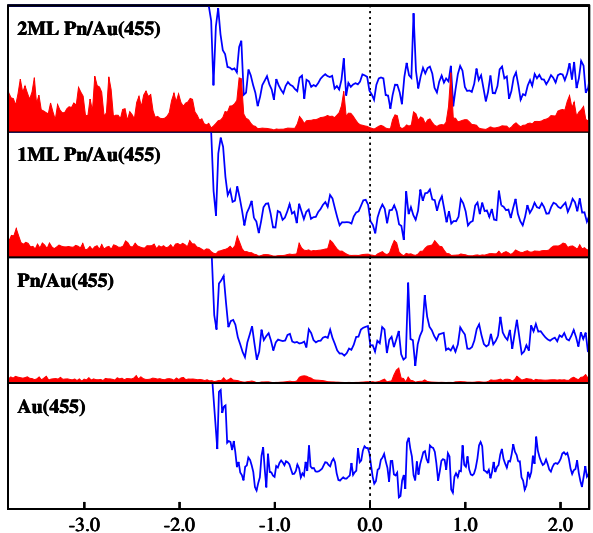} 
\caption{The projected DOS structures of Pen on the vicinal 
Au(455) surface calculated at the optB86b-vdW level of theory. \label{fig8}}
\end{figure}

The PBE XC functional follows the same trend and optimizes 2 ML 
Pen into tilted-tilted configuration (Figure~\ref{fig7}c).  The interaction 
between the Pen films and the gold surface estimated by the standard 
DFT is weaker than what is experimentally observed. This shortcoming
of LDA parametrization of long range correlations is also implied by 
the calculated layer heights from the surface being unreasonably large.
For instance, for the 1 ML case, PBE results in a Pen layer which stays 
3.92 {\AA} above the surface in the tilted geometry. This distance is 
predicted to be 3.18 {\AA} in the flat2 model relaxed using the 
optB86b-vdW approach. Moreover, at the PBE level, the presence of 
the step does not seem to cause significant disorder in the formation 
of thin pentacene films on gold. On the contrary, tilted Pentacene at 
the step edge acts as a mediator of molecular tilting throughout the 
first and the second adlayers. Indeed, when we start with the flat2-tilted 
initial configuration, the geometry optimization with PBE XC functional 
resulted in the tilted-tilted model structure as indicated in Table~\ref{table4}.
On the other hand, the vdW-DF theory of the van der Waals interactions 
leads to a relatively less ordered flat2-tilted formation (Figure~\ref{fig7}d). 
Furthermore, optB86b-vdW relaxes the tilted-tilted configuration, which is 
favored by PBE, to flat2-tilted model as the minimum energy structure. 
When one compares the 2 ML Pen adsorption results on the flat Au(111) and 
on the vicinal Au(455) surface,  the effect of the step becomes vital in
the formation of thin film formation. Especially, vdW corrected DFT 
indicates disorder to a certain extent in the vicinity of the step edge.

Electronically, we computed the DOS structures of the thin Pen films
up to 2 ML coverage using the optB86b-vdW approach. The densities 
of states of the bare and a single Pen adsorbed Au(455) surfaces are 
depicted in the bottom two panels of Figure~\ref{fig8}.  An isolated 
Pen at the step exhibits significant electronic contribution at and around
the Fermi energy as a result of the formation of four strong bonds with 
the Au atoms at the edge. The HOMO level of tilted Pen at the step
is identified as $-0.75$ eV while its LUMO lies 0.35 eV above the Fermi 
energy. Core molecular orbitals strongly resonate with the deeper lying 
gold states. 

The positions of the HOMO and the LUMO levels of Pen with respect to
the Fermi energy in the vicinal Au(455) surface case (Figure~\ref{fig8})  
are significantly red shifted relative to those in the flat Au(111) surface
case (Figure~\ref{fig4}). This level shift is approximately 0.4 eV.  
Moreover, Pen shows a larger dispersion of  molecular orbitals over 
Au(455) states. These results indicate an enhanced charge carrier transport 
at the Pen-gold interface in favor of the vicinal surface.

\section{Conclusions}
We have performed density functional theory calculation of pentacene films 
with and without dispersion corrections and investigated the evolution of 
crystal and electronic structure of the films as a function of coverage and 
surface steps. To study the effects of dispersion interactions we have used 
a self-consistent implementation of the vdW-DF method of Dion \textit{et al.} 
as implemented in VASP and compared the results with those obtained by the
standard PBE functional. While for isolated molecules both methods yield the 
same adsorption site as the most preferred one for a lying down molecule, 
the binding strength was observed to be higher with dispersion correction 
as expected. The energy difference between the different binding sites were 
very small resulting in a smooth potential energy surface for an isolated 
pentacene on Au(111). Three different experimentally observed unit cell 
structures were investigated and (6x3) was found to be the most stable one, 
though the binding energies for all three were very close to each other, the 
difference being less than 8 kJ/mol. These results in fact explain why 
pentacene has so many monolayer phases/structures forming simultaneously on 
the Au(111) surface. While dispersion corrected results favor a flat lying 
first layer for all the coverages on the (111) surface, the PBE results 
indicate a tilting of the molecules about their long axis which is not more 
than 10 degrees. These results are also in agreement with the experimental 
findings regarding pentacene monolayer films on Au(111) which report the 
molecules to be either flat or to have a very small tilt angle. In case of 
the vicinal (455) surface, an isolated pentacene molecule was found to be 
strongly bound to the step edge which also causes significant contribution 
to the density of states at and around the Fermi energy. Dispersion corrected 
projected DOS calculations indicate that for the multilayer films pentacene 
states are significantly shifted and dispersed on the (455) surface when 
compared with the (111) surface, which can result in an enhanced charge 
transfer at the Pen-gold interface. In summary, our results underline the 
importance of the dispersion corrections for the loosely bound systems like 
pentacene on gold and the role played by step edges in determining the 
multilayer film structure and charge transfer at the organic molecule-metal 
interface.


\end{document}